\documentclass[sigconf]{acmart}

\copyrightyear{2026}
\acmYear{2026}
\setcopyright{cc}
\setcctype{by}
\acmConference[SIGIR-AP 2026]{Proceedings of the 2026 Annual International ACM SIGIR Conference on Research and Development in Information Retrieval in the Asia Pacific Region}{December 13--15, 2026}{Singapore, Singapore}
\acmBooktitle{Proceedings of the 2026 Annual International ACM SIGIR Conference on Research and Development in Information Retrieval in the Asia Pacific Region (SIGIR-AP 2026), December 13--15, 2026, Singapore, Singapore}
\acmDOI{10.1145/3848000.3850212}
\acmISBN{979-8-4007-3033-7/2026/12}

\usepackage{balance}

\begin{document}

\title{Collaborative Search Meets Large Language Models: An Empirical Study}

\author{Peiyuan Gong}
\affiliation{%
  \institution{GSAI, Renmin University of China}
  \city{Beijing}
  \country{China}
}
\email{pygongnlp@gmail.com}

\author{Jiamian Li}
\affiliation{%
  \institution{GSAI, Renmin University of China}
  \city{Beijing}
  \country{China}
}
\email{miankora33@ruc.edu.cn}

\author{Erhan Zhang}
\affiliation{%
  \institution{GSAI, Renmin University of China}
  \city{Beijing}
  \country{China}
}
\email{erhanzhang@ruc.edu.cn}

\author{Fengming Liang}
\affiliation{%
  \institution{GSAI, Renmin University of China}
  \city{Beijing}
  \country{China}
}
\email{05liangfm@gmail.com}

\author{Kexin Liu}
\affiliation{%
  \institution{GSAI, Renmin University of China}
  \city{Beijing}
  \country{China}
}
\email{minnieliu6688@gmail.com}

\author{Yue Chen}
\affiliation{%
  \institution{GSAI, Renmin University of China}
  \city{Beijing}
  \country{China}
}
\email{2022201395@ruc.edu.cn}

\author{Jiaxin Mao}
\authornote{Corresponding author.}
\affiliation{%
  \institution{GSAI, Renmin University of China}
  \city{Beijing}
  \country{China}
}
\email{maojiaxin@gmail.com}

\renewcommand{\shortauthors}{Peiyuan Gong et al.}

\begin{abstract}
Collaborative search enables users to address complex information needs jointly. Advances in large language models (LLMs) enable autonomous agents to support collaborative information seeking within shared communication spaces. 
However, little is known about how these agents shape search experiences, behaviors, and outcomes compared with conventional search tools. 
We conduct a user study with 80 students working in 40 pairs. 
Each pair completes four collaborative search tasks under four scenarios: QAAgent, which answers queries using an LLM’s parametric knowledge; CoSearchAgent, which uses multi-user conversational context to reformulate queries, retrieve web information, and generate source-grounded answers, while asking clarifying questions when needed; an in-chat search plugin that returns result lists; and manual Google Search. 
After each task, participants complete several questionnaires assessing their search experience and independently write a report; after all tasks, they participate in individual interviews.
Results show that LLM-based agents reduce perceived workload, support collaboration, and reshape communication and evidence-use practices. 
They also encourage more independent paraphrasing, albeit at the cost of report richness.
To the best of our knowledge, this is the first systematic empirical comparison of LLM-based agents and conventional tools for collaborative search across search experience, user behavior, and collaborative outcomes. 
\end{abstract}



\maketitle

\section{Introduction}
Collaborative search enables users to jointly address complex information needs by searching for, evaluating, and sharing information \cite{shah2010collaborative, shah2014collaborative}. 
Compared with individual search, it allows collaborators to pool complementary knowledge, refine understanding through discussion, and synthesize diverse perspectives into a shared solution \cite{amershi2008cosearch, putra2018searchx, gonzalez2011coagmento}. 
To support these activities, collaborative search systems typically provide shared workspaces, real-time communication, query histories, and tools for organizing and annotating information. 
Prior research suggests that collaborative search is widespread, with many users engaging in it at least weekly \cite{morris2013collaborative}.
These findings underscore the importance of systems that support both information seeking and collaboration.

\textbf{\begin{figure*}[t] 
  \centering
  \includegraphics[width=0.9\linewidth]{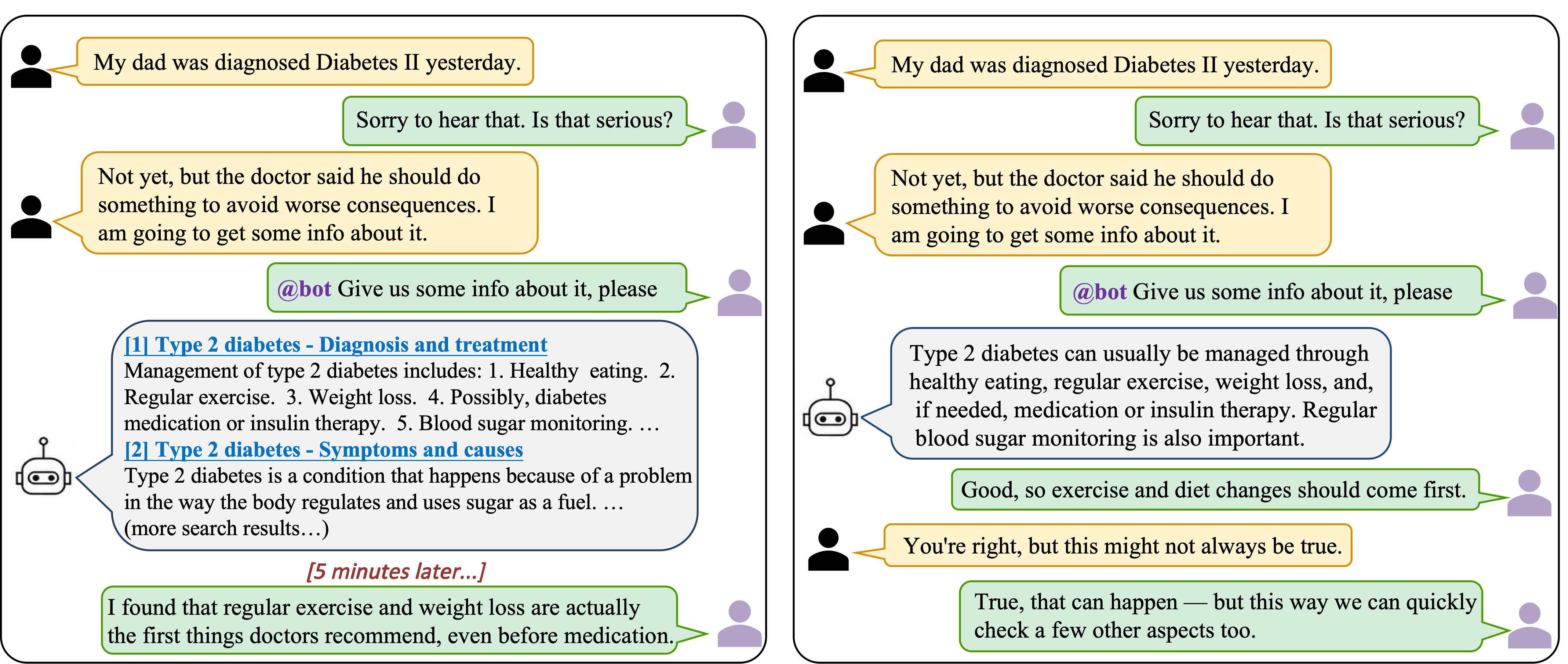}
  \caption{Two lightweight collaborative search systems: (I) Search Plugin, which provides reliable sources but requires longer browsing; and (II) QAAgent, which offers convenient answers that speed up discussion but may leave users unsure.}
  \label{intro}
\end{figure*}}

Early collaborative search systems typically took the form of dedicated standalone applications \cite{morris2007searchtogether, paul2009cosense, gonzalez2011coagmento, capra2012resultsspace, putra2018searchx, diriye2012querium, amershi2008cosearch}. 
They commonly integrated communication and search interfaces, providing features such as shared query histories \cite{morris2007searchtogether, paul2009cosense, putra2018searchx, capra2012resultsspace, diriye2012querium, gonzalez2011coagmento}, group notes and summaries \cite{amershi2008cosearch, paul2009cosense, gonzalez2011coagmento, diriye2012querium}, and mechanisms for recommending or sharing results \cite{morris2007searchtogether, gonzalez2011coagmento, diriye2012querium}. 
However, users often prefer coordinating through familiar instant messaging platforms while accessing conventional search tools separately because these platforms are familiar, convenient, and ubiquitous \cite{morris2013collaborative}. 
This preference motivates lightweight collaborative search systems that embed search functionality in instant messaging platforms and present results within multi-user conversations \cite{avula2018searchbots, avula2019embedding, avula2020wizard, avula2022effects}. 
Presenting retrieved information within the conversational interface reduces application switching and lets collaborators discuss and use results as part of their ongoing discourse.

Recent advances in large language models (LLMs) create opportunities to extend lightweight collaborative search beyond retrieving and presenting results.
LLMs can interpret evolving conversational context, infer user intent \cite{li2023camel, chen2023felm, qian2024tell}, and interact through natural language \cite{gong2024cosearchagent, tan2023chatgpt, mao2024multi}.
These capabilities are valuable in multi-user settings, where information needs may span messages, evolve through discussion, or remain underspecified.
By synthesizing conversations, LLMs can reformulate users' information needs into effective queries, identify missing or ambiguous details, and ask clarifying questions before initiating or refining searches.
Augmented with web retrieval, LLMs can synthesize source-grounded information, reduce unsupported claims, and cite supporting sources \cite{jiang2023active, yan2024corrective, jin2025flashrag}.
More broadly, LLMs may automate adaptive search assistance often implemented through the Wizard-of-Oz paradigm \cite{avula2018searchbots, avula2019embedding, avula2020wizard, avula2022effects, avula2023and}, making it more feasible to deploy and evaluate in real-world collaborative environments.

However, little research has systematically compared LLM agents with traditional search tools in collaborative search. 
It remains unclear how their richer, adaptive assistance affects search experience, shapes user behavior, and influences collaborative outcomes.
Figure~\ref{intro} contrasts a traditional search plugin with a QA agent. 
The search plugin delivers timely, source-grounded information, but may interrupt conversations and require users to evaluate and integrate results.
The QA agent can synthesize information, adapt to context, and offer proactive support, yet may hallucinate or reduce engagement with original sources. 
These differences may affect how users formulate queries, verify and share information, divide tasks, coordinate decisions, and perceive workload and satisfaction. 
A systematic comparison is therefore needed across LLM agents, embedded search plugins, and standalone search engines.

To address this gap, we conduct a user study with 80 students organized into 40 pairs, systematically comparing two forms of LLM-based assistance with two conventional search tools. 
Specifically, QAAgent answers users' questions directly using the LLM's parametric knowledge, whereas CoSearchAgent~\cite{gong2024cosearchagent} interprets the multi-user conversational context, rewrites queries, proactively asks clarifying questions, retrieves information from the web, and generates source-grounded answers. 
We compare these agents with two conventional search tools: 
SearchAgent, a lightweight plugin that returns ranked result lists within the group conversation, 
and manual Google search, through which collaborators search independently and share their findings in the conversation. 
Each pair completes four open-ended collaborative search tasks, one under each scenario.
After each task, each participant completes several questionnaires assessing their search experience and then writes a brief search report.
After completing all four tasks, they participate in individual interviews.

Our results reveal that LLM-based assistance reshapes collaborative search by reducing perceived task demands and strengthening teamwork.
QAAgent delivers the broadest collaborative gains, improving mutual awareness, information sharing, and exchange efficiency.
CoSearchAgent provides the best agent experience and reduces evidence inspection through synthesized, web-grounded answers.
However, these experiential benefits do not necessarily improve report quality. In the plug-in setting with a fixed word limit, reports assembled from copied webpage content are of higher quality than those based on copied agent responses, as webpages provide richer information for selection and synthesis.
This helps explain why Google Search and SearchAgent produce richer reports, whereas LLM-based agents encourage independent wording.
Overall, automated retrieval and synthesis reduce effort, but latency and clarification can offset time savings, while direct answers raise concerns about completeness and trust.
Effective agents should organize shared knowledge, connect evidence, guide task allocation, and keep proactive assistance under user control.
The code, study scripts, and demo are available in our \href{https://github.com/pygongnlp/CoSearchAgent}{repository}.

This work makes three contributions:
\begin{itemize}
    \item To our knowledge, this is the first systematic comparison of LLM-based agents and traditional tools for collaborative search, focusing on search experience, user behavior, and collaborative outcomes.
    
    \item We study 80 participants in 40 pairs, each completing four tasks using two LLM-based systems, a conversational search plugin, and Google Search.
    
    \item We reveal how LLM assistance reshapes search and collaboration, identifying key trade-offs and design implications for knowledge organization, verification, exploration, and proactive support.
\end{itemize}

\section{Related Work}
In this section, we review related work on collaborative search from two perspectives: system design and user behavior.

\subsection{System Construction}
Early collaborative search systems are dedicated applications integrating search with coordination support: CoSearch and SearchTogether support querying, communication, and activity awareness \cite{amershi2008cosearch, morris2007searchtogether}; CoSense and Coagmento offer workspaces and annotations \cite{paul2009cosense, gonzalez2011coagmento}; and ResultsSpace, Querium, and SearchX support recommendations and synthesis \cite{capra2012resultsspace, diriye2012querium, putra2018searchx}. Collaborative ranking and allocation use interaction histories and relevance feedback to prioritize and distribute results among group members \cite{freyne2004further, pickens2008algorithmic, soulier2013collaborative}. Despite reducing duplication and supporting evidence integration, dedicated systems require users to adopt unfamiliar interfaces and workflows. Users’ reliance on familiar communication tools and search engines \cite{morris2013collaborative} motivates lightweight, embedded support. Slack-based searchbots enable information requests, sharing, and discussion within ongoing group conversations \cite{avula2018searchbots, avula2019embedding}. Wizard-of-Oz studies extend these bots toward proactive guidance through clarification, suggestions, and dialogue- and task-level initiative \cite{avula2020wizard, avula2022effects, avula2023and}. LLMs further automate and integrate these capabilities into conversational workflows: CoSearchAgent interprets multi-user conversations, retrieves web information, generates grounded answers, and asks clarification questions \cite{gong2024cosearchagent}, enabling agents to participate in and guide collaborative search.

\subsection{User Behavior Analysis}
Collaborative search brings together searching, communication, and teamwork. 
Group members explain what they need, divide tasks, follow each other’s progress, and combine the evidence they find \cite{hertzum2008collaborative, shah2010awareness}. 
Research in this area looks at how groups break down tasks, revise queries, discuss which results are useful, and bring information together \cite{shah2010collaborative, shah2014collaborative, hertzum2019empirical}. 
Shared search histories and workspaces, along with cues about others’ activity, help members divide tasks and avoid repeating the same searches. 
Systems can also assign results based on each member’s role or expertise \cite{morris2007searchtogether, pickens2008algorithmic, shah2010awareness}. Communication helps groups coordinate, but too little, too much, or poorly timed communication can make coordination harder \cite{gonzalez2013let}. At the system level, conversational search lets users ask for information, discuss evidence, and involve agents in the search \cite{avula2018searchbots, avula2019embedding}. Proactive agents can also clarify what users need and suggest where to search next \cite{aliannejadi2021analysing, avula2022effects, avula2023and}. However, it is still unclear whether LLM-based agents work better than traditional tools. Many studies examine one feature or design at a time, rather than its combined effects on search experience, user behavior, and group results. To address this gap, we compare QAAgent, CoSearchAgent, SearchAgent, and Google Search in terms of workload, collaboration, communication, evidence use, and report quality.

\section{LLM-Based Collaborative Search Agents}
\label{llm_based_collaborative_search_agents}
In this section, we introduce two LLM-based agents for collaborative search: QAAgent and CoSearchAgent \cite{gong2024cosearchagent}. 

\subsection{QAAgent}
\label{qaagent_method}
We first propose QAAgent, a question-answering plugin that draws on an LLM’s pretrained knowledge \cite{roberts2020much, petroni2019language}. 
Specifically, given a user query $q$, we construct a prompt $P_{qa}$ to guide the LLM in generating a relevant answer:

\begin{equation}
    a_{qa} = P_{qa}(LLM, q)
\end{equation}

where $a_{qa}$ is generated directly by the LLM.

\subsection{CoSearchAgent}
\label{cosearchagent_method}
We reimplement CoSearchAgent~\cite{gong2024cosearchagent} to evaluate more sophisticated LLM-based support for collaborative search. It formulates search queries from ongoing discussions, retrieves relevant web evidence, and synthesizes this evidence into answers.

\subsubsection{Conversational Query Understanding}
Requests made during a conversation are often incomplete or context-dependent~\cite {ye2023enhancing,mo2023convgqr}. 
This issue is especially important in collaborative search because a user's request may refer to information introduced earlier by another group member. 
Let $U=\{u_1,u_2,\ldots,u_n\}$ denote the multi-user conversation history and $q$ the current user query.
We first use a query-rewriting prompt $P_{\mathrm{rewrite}}$ to resolve references and recover omitted context from the conversation, producing a self-contained query for web search:

\begin{equation}
    (q_{\mathrm{rewrite}},o)=P_{\mathrm{rewrite}}(LLM,U,q),
\end{equation}

where $q_{\mathrm{rewrite}}$ is the rewritten query ($q_{\mathrm{rewrite}}=q$ if $q$ is self-contained), and $o$ indicates whether clarification is needed.

When an ambiguity cannot be resolved from the conversation history, we use a clarification prompt $P_{\mathrm{clarify}}$ to generate a clarifying question:

\begin{equation}
    q_{\mathrm{clarify}}=P_{\mathrm{clarify}}(LLM,U,q_{\mathrm{rewrite}}).
\end{equation}

After the user responds, CoSearchAgent incorporates the response into the conversation history and reassesses the query. 
It initiates a web search once no further clarification is required.

\subsubsection{Web Evidence Extraction}

CoSearchAgent submits the rewritten query $q_{\mathrm{rewrite}}$ to a search engine and obtains the top-$m$ results:

\begin{equation}
    SERP=Search(q_{\mathrm{rewrite}})=\{(t_i,l_i,s_i)\}_{i=1}^{m},
\end{equation}

where $t_i$, $l_i$, and $s_i$ denote the title, link, and search-engine snippet of result $i$, respectively.

For each result, the agent fetches the page content $h_i$ and uses an evidence-extraction prompt $P_{\mathrm{extract}}$ to identify evidence relevant to the rewritten query:

\begin{equation}
    ref_i=P_{\mathrm{extract}}(LLM,q_{\mathrm{rewrite}},h_i).
\end{equation}

The agent then presents the retrieved sources with the extracted evidence in place of the original snippets:

\begin{equation}
    SERP_c=\{(t_i,l_i,ref_i)\}_{i=1}^{m}.
\end{equation}

This representation retains access to the original sources while foregrounding the passages most relevant to the group's information need.

\subsubsection{Source-Grounded Answer Generation}

Given the rewritten query $q_{\mathrm{rewrite}}$ and extracted evidence $REF=\{ref_1,ref_2,\ldots,ref_m\}$, CoSearchAgent uses a retrieval-augmented generation prompt $P_{\mathrm{rag}}$ to synthesize a source-grounded answer:

\begin{equation}
    a_c=P_{\mathrm{rag}}(LLM,q_{\mathrm{rewrite}},REF).
\end{equation}

The answer $a_c$ includes citations that link its claims to the supporting sources.
CoSearchAgent thus delegates query interpretation, retrieval, and answer synthesis while preserving inspectable evidence for verification and discussion.

\begin{figure*}
    \centering
    \includegraphics[width=0.9\linewidth]{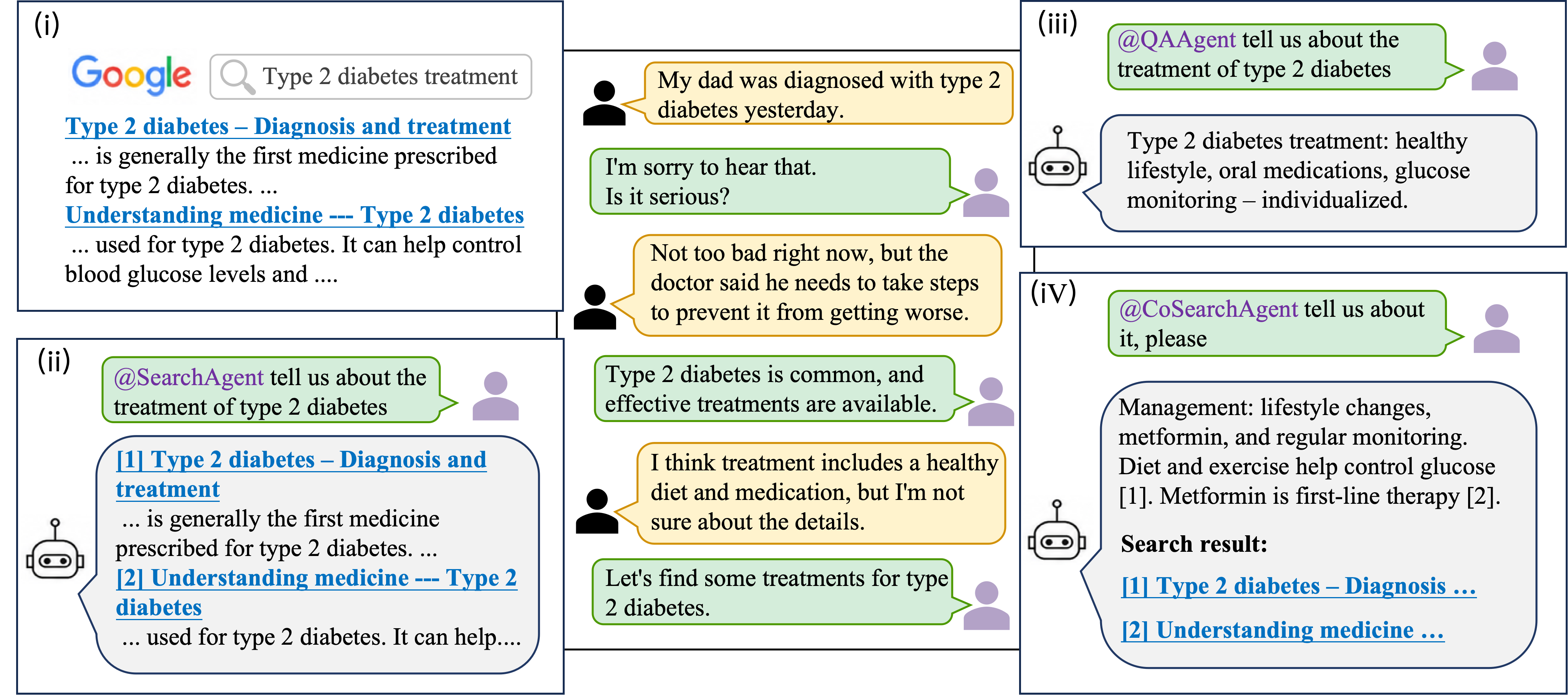}
    \caption{
Four search scenarios in the user study: (I) Google Search, used independently outside the conversation; (II) SearchAgent, integrating results into the conversation; (III) QAAgent, directly answering queries; and (IV) CoSearchAgent, interpreting multi-user conversations, retrieving relevant information, and providing tailored responses.}
    \label{agents}
\end{figure*}

\section{User Study}
\label{sec:user-study}
To evaluate the effectiveness of LLM-based agents in collaborative search, we conduct a user study with 80 students organized into 40 pairs. 
The following subsections describe the research questions (Section~\ref{rqs}), experimental procedure (Section~\ref{sec:study-procedure}), tasks (Section~\ref{sec:search-tasks}), search scenarios (Section~\ref{sec:search-scenarios}), questionnaires (Section~\ref{sec:study-questionnaires}), and face-to-face interviews (Section~\ref{sec:study-interview}).

\subsection{Research Questions}
\label{rqs}
We investigate how the use of LLM-based agents, as opposed to traditional search tools, influences collaborative search across three key dimensions:

\noindent\textbf{RQ1.} What differences in participants' search experiences arise between LLM-based agent and traditional search scenarios?

\noindent\textbf{RQ2.} In what ways does the choice of search tools shape participants' search strategies and collaboration behaviors?

\noindent\textbf{RQ3.} To what extent does the use of LLM-based agents influence the quality of collaboratively produced final reports?

\begin{table*}[t]
  \caption{Collaborative search tasks.}
  \label{tab:search-tasks}
  \centering
  \small
  \setlength{\tabcolsep}{5pt}
  \renewcommand{\arraystretch}{1.08}
  \begin{tabular}{p{0.16\textwidth} p{0.42\textwidth} p{0.36\textwidth}}
    \toprule
    \textbf{Task} & \textbf{Background} & \textbf{Goal} \\
    \midrule
    AI painting & Two students explore how artificial intelligence is used in painting for an introductory course. & Investigate AI painting collaboratively and write individual learning reports. \\
    Childhood autism & Two volunteers help parents seeking information about possible autism in their two-year-old child. & Investigate childhood autism collaboratively and write individual learning reports. \\
    Romantic music & Two concertgoers want to understand Western music from the Romantic period before attending a performance. & Investigate Romantic-period music collaboratively and write individual learning reports. \\
    Air pollution & Two contestants prepare for an environmental knowledge competition concerning air pollution. & Investigate air pollution collaboratively and write individual learning reports. \\
    \bottomrule
  \end{tabular}
\end{table*}

\subsection{Procedure}
\label{sec:study-procedure}
We recruit 80 students and form 40 participant pairs. 
Each pair completes four collaborative search tasks, one under each search scenario: Google Search, SearchAgent, QAAgent, and CoSearchAgent.
We use two orthogonal \(4 \times 4\) Latin squares to counterbalance task–scenario assignments and presentation order \citep{bradley1958complete,reese1997counterbalancing}. 
Their superposition yields four base sequences that collectively cover all 16 task–scenario combinations. 
Four cyclic rotations of each base sequence produce 16 variants, ensuring that each combination appears once in each ordinal position. 
We assign 10 pairs to each base-sequence family and distribute them as evenly as possible across its four rotations. Thus, each pair experiences every search scenario once, and each task–scenario combination is completed by 10 pairs.

\textbf{[Instructions]} Before the experiment, moderators explain the procedure and demonstrate each search scenario. 
Pair members then briefly introduce themselves. 
For each task, pairs have up to 20 minutes to search, discuss, divide the work, develop a shared understanding, and plan their learning reports. 
Afterward, each participant independently writes a learning report of at least 500 words in Tencent Docs within 10 minutes and completes a post-task questionnaire. 
A break of up to 5 minutes is allowed between tasks. After all four tasks, each participant attends an individual face-to-face interview.

\textbf{[Tools]} The study uses two moderators, two computer workstations, and three rooms.
The moderators conduct the briefing in a common room, after which the two participants of each pair work at separate computers in separate rooms. 
Slack provides a dedicated channel for communication and access to the assigned agents.
Tencent Meeting records the sessions, Google Chrome supports web access when permitted by the assigned search scenario, and Tencent Docs supports the preparation and submission of individual learning reports.

\textbf{[Limits]} Participants must communicate with their partner only through the dedicated Slack channel and may not communicate verbally at any time during the tasks.
They may not open software other than the specified experimental tools.
Information seeking must follow the assigned search scenario: participants may search independently only in the Google Search scenario and may not switch to another search tool in the three agent scenarios.
Although participants may discuss report structure during the collaborative session, each participant must write and submit the final learning report independently.

\subsection{Tasks}
\label{sec:search-tasks}
In our user study, each pair completes four open-ended knowledge-exploration tasks covering AI painting, childhood autism, Romantic-period Western music, and air pollution. Each task presents a background scenario that establishes the participants' roles and information needs, followed by a goal that requires the pair to investigate the topic collaboratively and each participant to submit an individual learning report. The tasks have no single correct answer and are designed to encourage participants to decompose their information needs, divide subtopics, exchange evidence, and synthesize what they learn. Table~\ref{tab:search-tasks} summarizes the four tasks.

\subsection{Scenarios}
\label{sec:search-scenarios}
Each collaborative search task is paired with one of four search scenarios. All scenarios use Slack as the shared communication environment but differ in where participants search, how queries are formulated, and whether the returned information is synthesized by an LLM. The following three agents are embedded in Slack and can be invoked by typing ``@name.'' Figure~\ref{agents} illustrates the four scenarios, which are presented to participants as follows:

\begin{itemize}
  \item \textbf{Google Search.} Participants leave the Slack conversation to search independently with Google. They inspect the results themselves and return to Slack to share selected links, excerpts, or findings with their partner.
  \item \textbf{SearchAgent.} Participants invoke SearchAgent within Slack. The agent retrieves and returns a list of web search results for participants to inspect and integrate into their ongoing conversation; it does not synthesize a direct answer.
  \item \textbf{QAAgent.} Participants invoke QAAgent within Slack and receive a direct answer generated from the LLM's parametric knowledge. The agent does not retrieve web results or provide a search-result list.
  \item \textbf{CoSearchAgent.} Participants invoke CoSearchAgent within Slack. The agent interprets the multi-user conversation, rewrites user queries, retrieves relevant web information, and returns a context-aware response with inspectable results and source references, while proactively asking clarifying questions when needed.
\end{itemize}

\subsection{Questionnaires}
\label{sec:study-questionnaires}

We administer three seven-point (1 $\rightarrow$ 7) questionnaires assessing workload, collaboration experience, and agent experience. Participants complete the workload and collaboration-experience questionnaires after each task and the agent-experience questionnaire after each agent scenario. Questionnaires are detailed below:

\begin{itemize}
    \item \textbf{Workload.} Table~\ref{tab:workload-questionnaire} lists the six items administered after every task to assess mental demand, physical demand, temporal pressure, perceived performance, effort, and frustration.

    \item \textbf{Collaboration experience.} Table~\ref{tab:collaboration-questionnaire} presents the five items administered after every task to assess mutual awareness, information sharing, division of labor, and communication continuity.

    \item \textbf{Agent experience.} After each of the three agent scenarios, participants complete the seven items reported in Table~\ref{tab:agent-questionnaire}. These items assess information usefulness and completeness, time savings, workload reduction, support for collaboration, information reliability, and overall satisfaction.
\end{itemize}

\begin{figure*}[t]
  \centering
  \includegraphics[width=0.9\textwidth]{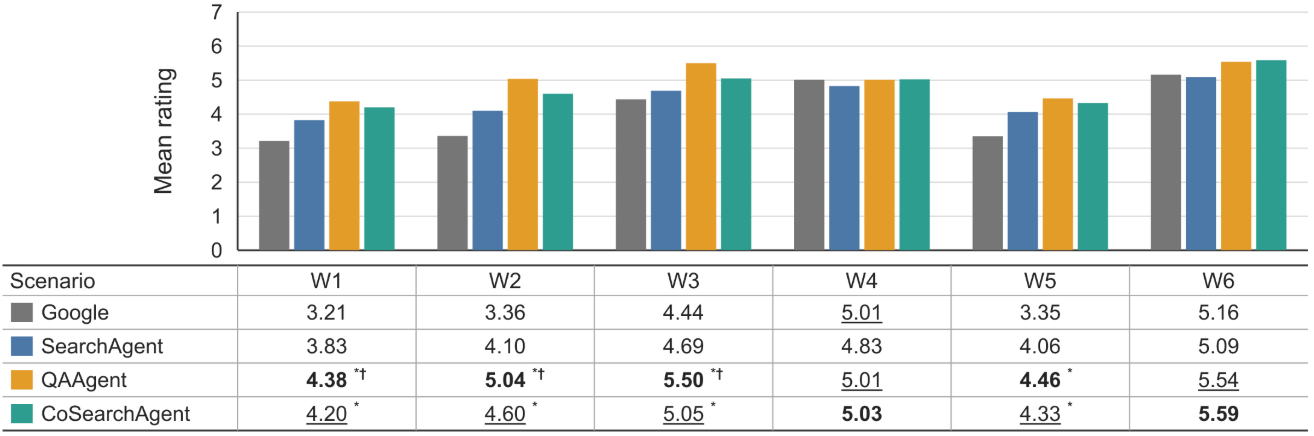}
  \caption{Mean workload ratings across scenarios. $^{*}$ and $^{\dagger}$ indicate significant differences from Google and SearchAgent, respectively ($p<.05$).}
  \label{fig:workload}
\end{figure*}

\begin{table}[t]
  \caption{Workload Questionnaire}
  \label{tab:workload-questionnaire}
  \centering
  \footnotesize
  \setlength{\tabcolsep}{3pt}
  \renewcommand{\arraystretch}{1.08}
  \begin{tabular}{@{}p{0.68\columnwidth} p{0.26\columnwidth}@{}}
    \toprule
    \textbf{Question} & \textbf{Score Range} \\
    \midrule
    {[W1]} How much mental activity, such as thinking, judging, decision-making, and searching, was required? & High $\rightarrow$ Low \\
    {[W2]} How much physical activity, such as clicking and browsing, was required? & High $\rightarrow$ Low \\
    {[W3]} How rushed or relaxed was the task pace? & Rushed $\rightarrow$ Relaxed \\
    {[W4]} How well did you complete the task? & Failure $\rightarrow$ Perfect \\
    {[W5]} How much effort did you expend? & Much $\rightarrow$ Little \\
    {[W6]} To what extent did you feel uneasy, discouraged, frustrated, or irritated? & Much $\rightarrow$ Little \\
    \bottomrule
  \end{tabular}
\end{table}

\begin{table}[t]
  \caption{Collaboration Experience Questionnaire}
  \label{tab:collaboration-questionnaire}
  \centering
  \footnotesize
  \setlength{\tabcolsep}{3pt}
  \renewcommand{\arraystretch}{1.08}
  \begin{tabular}{@{}p{0.68\columnwidth} p{0.26\columnwidth}@{}}
    \toprule
    \textbf{Question} & \textbf{Score Range} \\
    \midrule
    {[C1]} During the task, I knew what my partner was doing. & Disagree $\rightarrow$ Agree \\
    {[C2]} During the task, my partner knew what I was doing. & Disagree $\rightarrow$ Agree \\
    {[C3]} It was easy to share information with my partner. & Disagree $\rightarrow$ Agree \\
    {[C4]} It was easy to divide the work with my partner. & Disagree $\rightarrow$ Agree \\
    {[C5]} Communication with my partner remained smooth and uninterrupted. & Disagree $\rightarrow$ Agree \\
    \bottomrule
  \end{tabular}
\end{table}

\begin{table}[t]
  \caption{Agent Experience Questionnaire}
  \label{tab:agent-questionnaire}
  \centering
  \footnotesize
  \setlength{\tabcolsep}{3pt}
  \renewcommand{\arraystretch}{1.08}
  \begin{tabular}{@{}p{0.68\columnwidth} p{0.26\columnwidth}@{}}
    \toprule
    \textbf{Question} & \textbf{Score Range} \\
    \midrule
    {[A1]} The information provided by the agent was useful. & Disagree $\rightarrow$ Agree \\
    {[A2]} The information provided by the agent was comprehensive. & Disagree $\rightarrow$ Agree \\
    {[A3]} Overall, the agent saved us time. & Disagree $\rightarrow$ Agree \\
    {[A4]} Overall, the agent reduced our workload. & Disagree $\rightarrow$ Agree \\
    {[A5]} Overall, the agent facilitated our collaboration. & Disagree $\rightarrow$ Agree \\
    {[A6]} The information provided by the agent was reliable. & Disagree $\rightarrow$ Agree \\
    {[A7]} Overall, how satisfied were you with the agent? & Dissatisfied $\rightarrow$ Satisfied \\
    \bottomrule
  \end{tabular}
\end{table}
\subsection{Interview}
\label{sec:study-interview}

Once all four collaborative search tasks are complete, each participant takes part individually in a face-to-face, semi-structured interview. 
Through open-ended questions, the interviews elicit detailed reflections that fixed-response questionnaires cannot capture. 
Participants are asked to 
(1) rank the four search scenarios from most to least preferred and describe their experience with each; 
(2) discuss the advantages and disadvantages of QAAgent relative to the other scenarios; 
(3) discuss the advantages and disadvantages of CoSearchAgent relative to the other scenarios; 
and (4) suggest additional capabilities that an AI assistant should provide for collaborative multi-user search.

\begin{figure*}[t]
  \centering
  \includegraphics[width=0.9\textwidth]{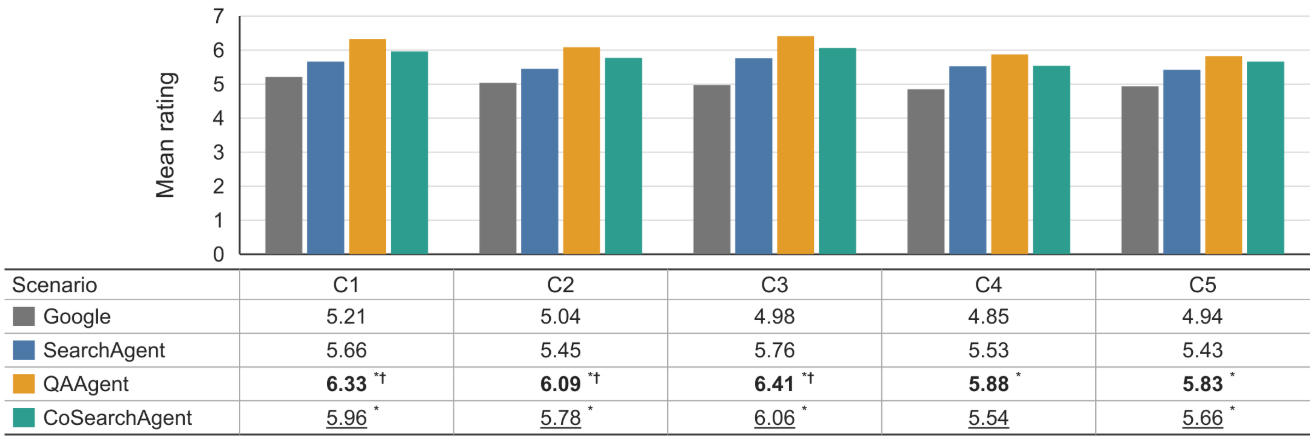}
  \caption{Mean collaboration-experience ratings across scenarios. $^{*}$ and $^{\dagger}$ indicate significant differences from Google and SearchAgent, respectively ($p<.05$).}
   \label{fig:collaboration-experience}
\end{figure*}

\begin{figure*}[t]
  \centering
  \includegraphics[width=0.9\textwidth]{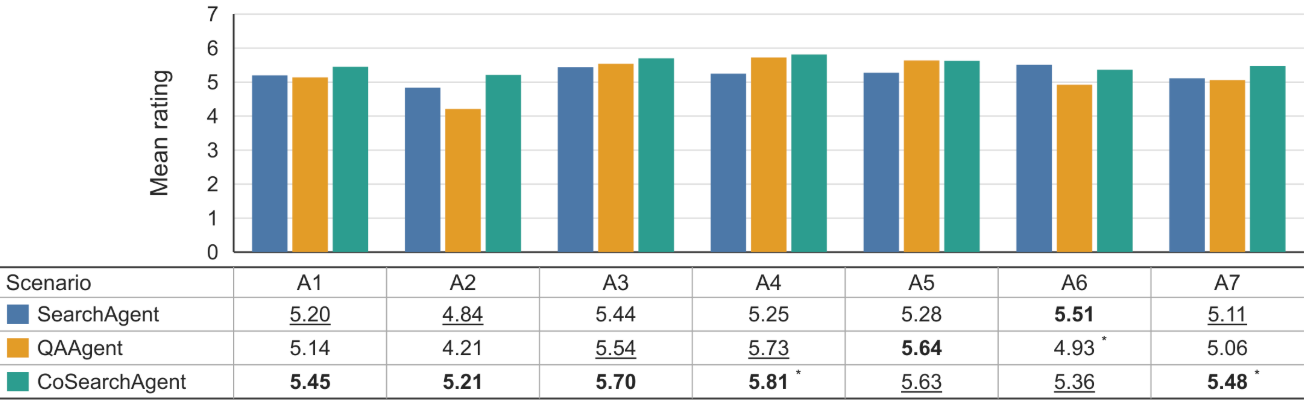}
    \caption{Mean agent-experience ratings across scenarios. $^{*}$ indicates a significant difference from SearchAgent ($p<.05$).}
  \label{fig:agent-experience}
\end{figure*}

\begin{figure}[t]
    \centering
    \includegraphics[width=\columnwidth]{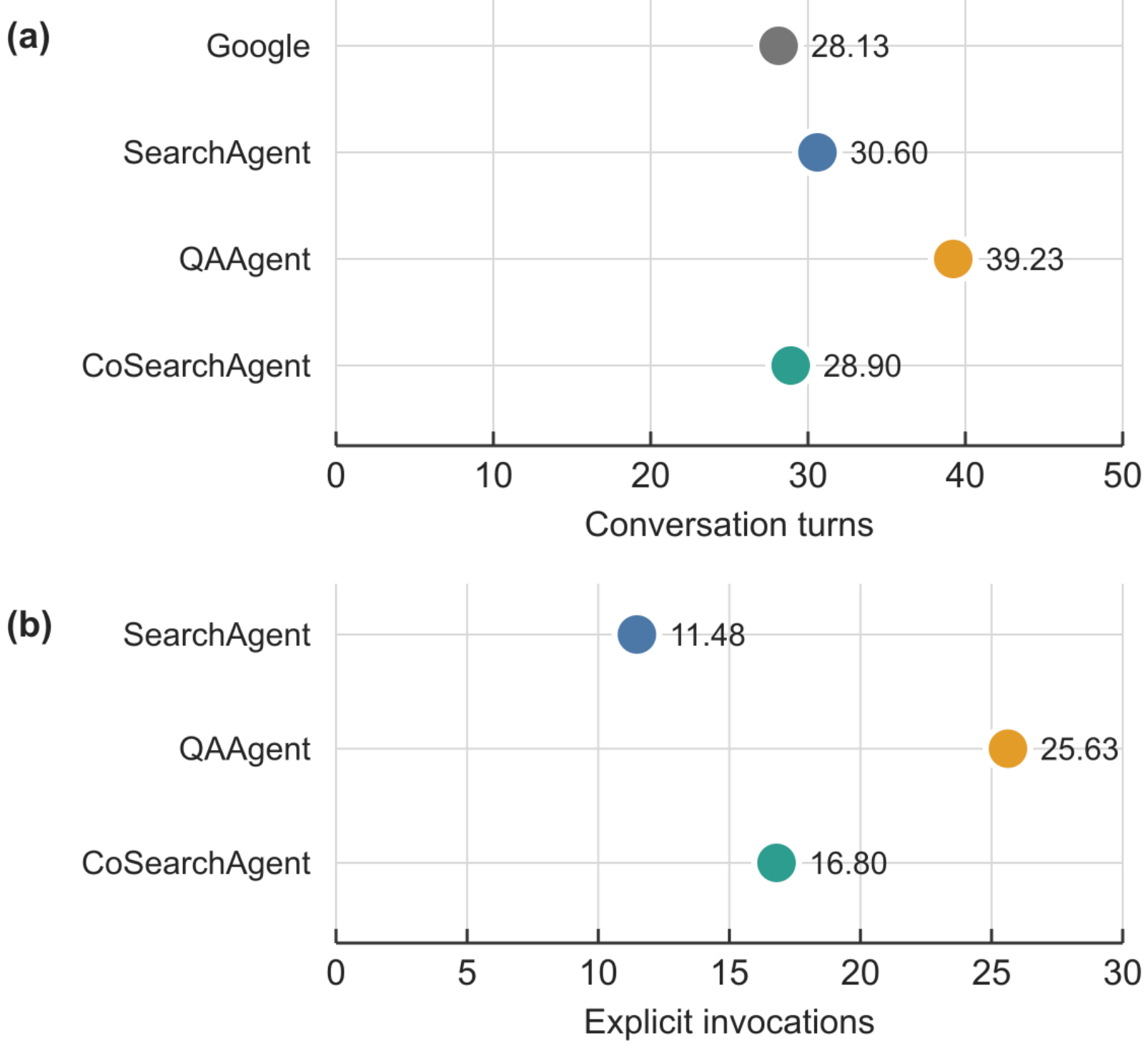}
    \caption{Communication behavior across scenarios: (a) conversation turns and (b) explicit agent invocations.}
    \label{fig:communication_behavior}
\end{figure}

\begin{figure}[t]
    \centering
    \includegraphics[width=\columnwidth]{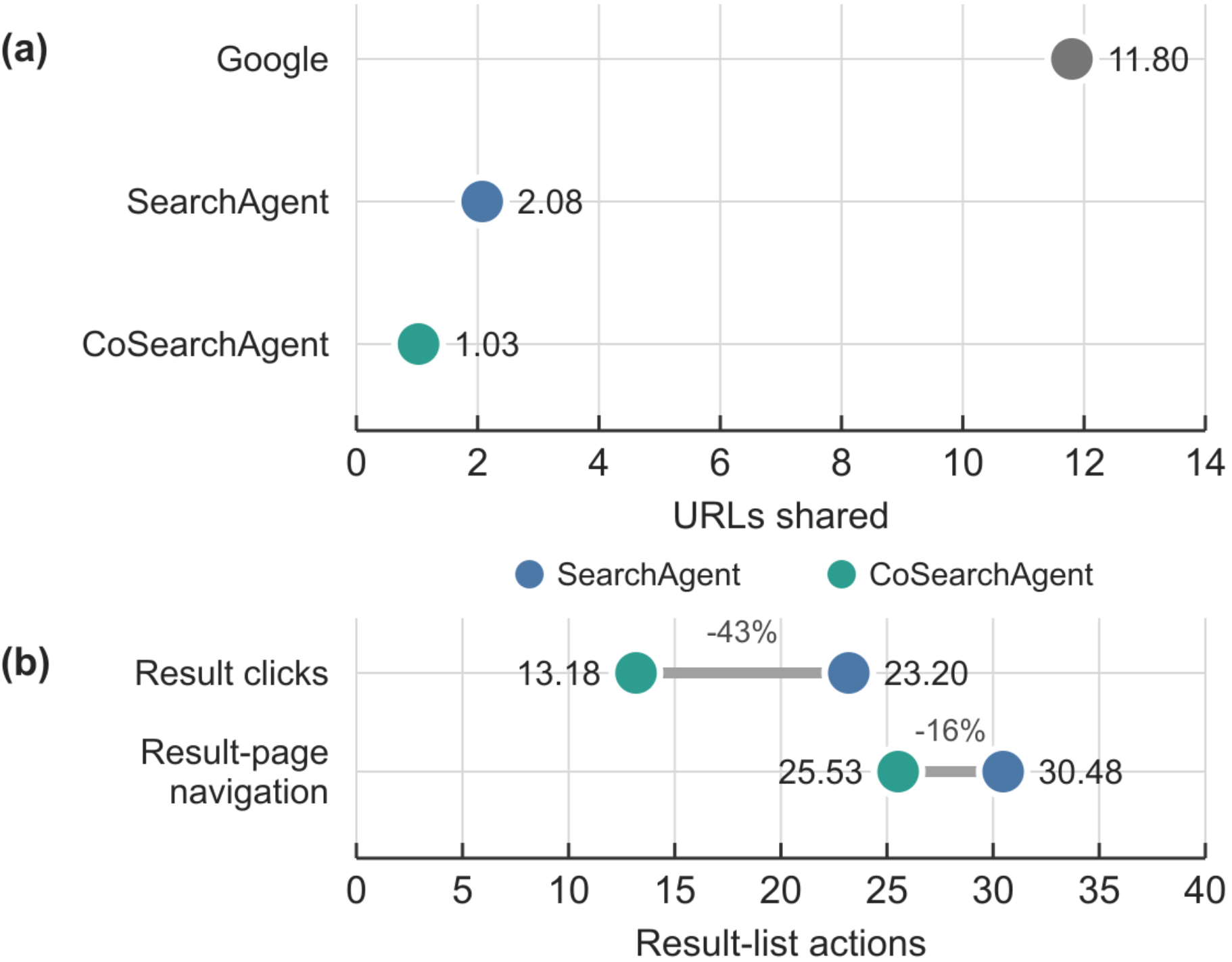}
    \caption{Evidence-use behavior: (a) URLs shared and (b) result-list actions; percentages show reductions from SearchAgent to CoSearchAgent.}
    \label{fig:evidence_use}
\end{figure}

\begin{figure}[t]
    \centering
    \includegraphics[width=\columnwidth]{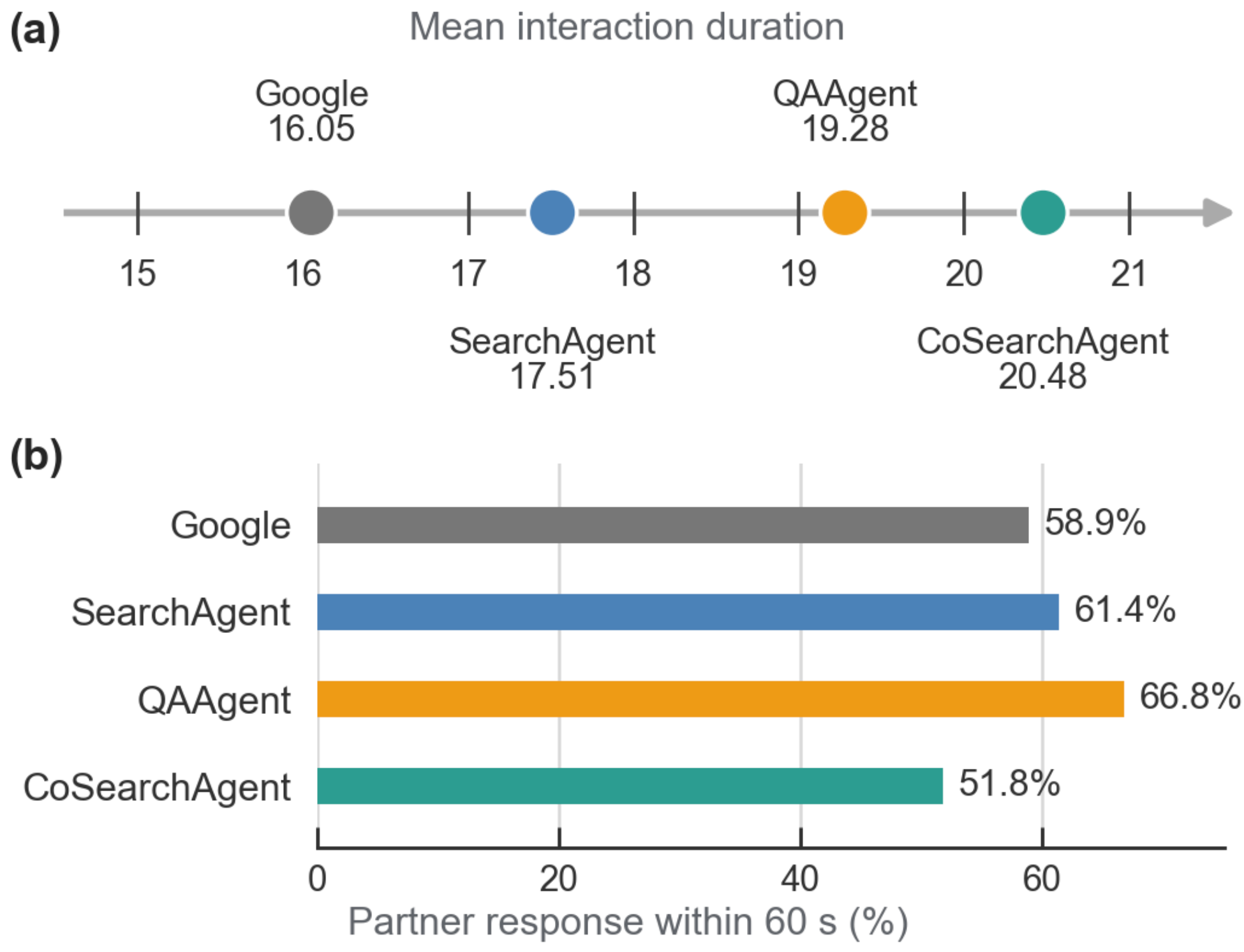}
    \caption{Temporal behavior across scenarios: (a) task-completion time and (b) 60-second partner-response rate.}
    \label{fig:temporal_behavior}
\end{figure}

\section{Results}
\label{sec:results}
We organize the results into three parts.
Section~\ref{sec:results-search-experience} examines participants' search experience through the workload, collaboration-experience, and agent-experience questionnaires. 
Section~\ref{sec:results-user-behavior} then checks user behavior through communication activity, evidence use, and task-completion time. 
Section~\ref{report_quality_sec} measures report qualities from various perspectives.

\subsection{Search Experience (RQ1)}
\label{sec:results-search-experience}

\subsubsection{Workload}
\label{sec:results-workload}
To evaluate workload, Figure~\ref{fig:workload} compares four scenarios across six dimensions, with higher scores indicating more favorable experiences. 
QAAgent achieves the highest ratings for mental demand (W1 = 4.38), physical demand (W2 = 5.04), temporal demand (W3 = 5.50), and effort (W5 = 4.46), whereas CoSearchAgent leads in perceived performance (W4 = 5.03) and frustration (W6 = 5.59). 
Both LLM-based agents significantly outperform Google Search on W1, W2, W3, and W5, while QAAgent outperforms SearchAgent on W1--W3. 
W4 and W6 show no significant pairwise differences, indicating that workload benefits of LLM-based assistance concern reduced task demands and effort rather than perceived performance or frustration.

\subsubsection{Collaboration Experience}
\label{sec:results-collaboration}
To assess collaboration experience, Figure~\ref{fig:collaboration-experience} compares mutual awareness, information sharing, division of labor, and communication continuity across the four scenarios. 
QAAgent receives the highest mean rating on all five items (5.83--6.41) and significantly outperforms Google Search on C1--C5, SearchAgent on C1--C3, and CoSearchAgent on C1. 
CoSearchAgent also significantly exceeds Google Search on C1, C2, C3, and C5. These results show that LLM-based assistance generally improves collaboration relative to conventional user-led search, with QAAgent producing the broadest gains, particularly in mutual awareness and information sharing.

\subsubsection{Agent Experience}
\label{sec:results-agent}
To examine participants' evaluations of the three systems, Figure~\ref{fig:agent-experience} compares usefulness, completeness, time savings, workload reduction, collaboration support, reliability, and overall satisfaction. 
CoSearchAgent ranks highest on five of seven items: usefulness (A1 = 5.45), completeness (A2 = 5.21), time savings (A3 = 5.70), workload reduction (A4 = 5.81), and overall satisfaction (A7 = 5.48). 
It significantly outperforms QAAgent on completeness, SearchAgent on workload reduction, and both agents on overall satisfaction.
QAAgent ranks highest on collaboration support (A5 = 5.64), while SearchAgent ranks highest on reliability (A6 = 5.51) and significantly outperforms QAAgent. Overall, CoSearchAgent provides the strongest experience, while QAAgent's lower completeness and reliability highlight the limitations of direct answer generation without web grounding.

\subsection{User Behavior (RQ2)}
\label{sec:results-user-behavior}

\subsubsection{Communication Activity}
To compare communication activity across search-support scenarios, we exclude agent-generated messages and analyze human-authored turns and agent invocations. 
Participants produce 39.23 turns per task with QAAgent, versus 30.60 with SearchAgent, 28.90 with CoSearchAgent, and 28.13 with Google Search (Figure~\ref{fig:communication_behavior}a). 
QAAgent is invoked most frequently (25.63 times per task), followed by CoSearchAgent (16.80) and SearchAgent (11.48) (Figure~\ref{fig:communication_behavior}b). 
Direct-answer assistance thus prompts repeated information requests and discussion, whereas communication under CoSearchAgent resembles that with conventional search tools.

\subsubsection{Evidence Use}
\label{sec:results-evidence}
To examine how search scenarios shape evidence use, we compare URL sharing and result-list interactions across scenarios with external sources. 
Participants share 11.80 URLs per task with Google Search, versus 2.08 with SearchAgent and 1.03 with CoSearchAgent (Figure~\ref{fig:evidence_use}a); 
QAAgent is excluded because it provides no external sources. 
This pattern suggests a shift from manually sharing sources found outside Slack to discussing evidence already available in the channel. 
Participants also click fewer results with CoSearchAgent than with SearchAgent (13.18 versus 23.20) and perform fewer result-page navigation actions (25.53 versus 30.48; Figure~\ref{fig:evidence_use}b). 
Thus, CoSearchAgent enables participants to begin with a synthesized, web-grounded answer and selectively inspect its sources for further detail or verification.

\subsubsection{Temporal Behavior}
\label{sec:results-time}
To measure how search support shapes the temporal organization of collaborative search, we assess interaction duration, defined as the time until partners cease exchanging messages, and the proportion of human-authored messages receiving a partner response within 60 seconds. 
Mean interaction duration is 16.05 minutes with Google Search, 17.51 with SearchAgent, 19.28 with QAAgent, and 20.48 with CoSearchAgent (Figure~\ref{fig:temporal_behavior}a). 
Agent-assisted scenarios therefore sustain interaction longer than Google Search. The 60-second response rate is highest for QAAgent (66.8\%), followed by SearchAgent (61.4\%), Google Search (58.9\%), and CoSearchAgent (51.8\%) (Figure~\ref{fig:temporal_behavior}b). 
QAAgent thus supports more rapid exchanges, whereas CoSearchAgent combines longer interaction with slower turn-taking, consistent with its clarification, retrieval, and synthesis stages.

\begin{table}[t]
\caption{Final-report quality. Bold: overall or web-evidence best; underline: second overall. $\dagger$: $p<.001$ vs. Google Search and SearchAgent.}
\label{tab:report-quality}
\centering
\small
\setlength{\tabcolsep}{3.5pt}
\renewcommand{\arraystretch}{1.08}
\begin{tabular*}{\columnwidth}{@{\extracolsep{\fill}}lcccc@{}}
\toprule
Scenario & Richness $\uparrow$ & Synthesis $\uparrow$ &
Independence $\uparrow$ & Gap $\downarrow$ \\
\midrule
Google Search & \textbf{4.56} & \textbf{4.64}
& 60.77\% & \underline{0.54} \\
SearchAgent & \underline{4.48} & \underline{4.62}
& 58.57\% & 0.67 \\
QAAgent & 4.21 & 4.48
& \textbf{99.45\%}$^{\dagger}$ & \textbf{0.50} \\
CoSearchAgent & 4.24 & 4.44
& \underline{83.84\%}$^{\dagger}$ & 0.64 \\
\bottomrule
\end{tabular*}
\end{table}

\subsection{Report Quality (RQ3)}
\label{report_quality_sec}
We assess 320 reports using four measures. Each report is evaluated twice by GPT-5.6-Sol using the task description and report text as input, with participant identities omitted.
\emph{Richness} (1--7) measures whether a report contains sufficient relevant and specific information. 
For example, a rich report on air pollution describes major pollutants, their sources, health effects, and solutions. 
\emph{Synthesis} (1--7) measures whether the report connects these facts into a clear explanation instead of simply listing them. 
\emph{Independence} is the percentage of 30-character text sequences that do not also appear in another pair's report on the same task; a higher value indicates less reuse of shared or source-like wording but does not by itself prove plagiarism or originality.
Finally, \emph{gap} is the absolute difference in overall quality scores between the two participants in the same pair, with a lower value indicating more consistent report quality.

Table~\ref{tab:report-quality} shows that Google Search and SearchAgent produce richer reports (4.56 and 4.48) than QAAgent and CoSearchAgent (4.21 and 4.24), partly because response-length limits on the lightweight agents in multi-party conversations make copied agent responses less informative than webpage content.
Independence exhibits the reverse pattern: QAAgent ranks highest (99.45\%), followed by CoSearchAgent (83.84\%), Google Search (60.77\%), and SearchAgent (58.57\%).
Both LLM-based agents significantly outperform the two conventional scenarios on this measure ($p<.001$).
Among the three scenarios that provide participants with access to browsable web evidence, CoSearchAgent records the highest independence score.
Pair gaps remain similar (0.50--0.67), with QAAgent exhibiting the smallest gap.
Overall, direct access to result pages helps participants include more factual detail but also leads to greater reuse of source wording.
QAAgent best supports independent expression, whereas CoSearchAgent provides the strongest balance between factual richness and independence.

\section{Discussion}
\label{sec:discussion}
We interpret quantitative results using post-study interviews.
We group participants' comments into recurring themes, and the number in parentheses indicates the number of unique participants whose responses are assigned to that theme.

\subsection{Agent Trade-offs}
\label{sec:discussion-agent-tradeoffs}

\subsubsection{Search Efficiency}
\label{sec:discussion-search-efficiency}
Compared with Google Search and SearchAgent, LLM-based agents reduce effort in locating and synthesizing information. 
QAAgent users value answers, efficiency (50), speed (16), and concision (14), matching favorable workload ratings. CoSearchAgent reduces manual inspection by combining synthesis with evidence. 
Yet lower effort does not shorten sessions. 
Google Search remains fastest, while users find CoSearchAgent slow (23) and its clarification process burdensome (13); QAAgent also demands prompt reformulation (17). 
Thus, LLM-based assistance improves efficiency by transferring search and synthesis to agents, but latency and interaction overhead limit time savings.

\subsubsection{Information Quality}
\label{sec:discussion-information-quality}
LLM-based agents simplify information consumption but raise completeness and trust concerns. 
These are strongest for QAAgent, which lacks source evidence: users report missing sources and low confidence (30), insufficient content (26), broad or generic explanations (20), and repetition (18). 
CoSearchAgent mitigates them by pairing summaries with inspectable links, valued by 52; 18 highlight reliability. 
Yet web grounding cannot guarantee quality: questionable sources (8), weakly related links (7), and irrelevant or misunderstood answers (11) persist. 
SearchAgent therefore retains the highest reliability rating, underscoring that effective LLM assistance requires both synthesis and reliable verification paths.

\subsubsection{Collaborative Interaction}
\label{sec:discussion-collaborative-interaction}
Embedding LLM-based agents in shared conversations strengthens collaboration because partners see queries and answers. QAAgent leads in mutual awareness, information sharing, work division, and communication continuity; use also turns direct answers into shared discussion resources. 
CoSearchAgent’s clarification questions (12) and context-sensitive interaction (9) help teams refine ambiguous needs. 
Yet benefits depend on accurate, economical interaction. 
CoSearchAgent may misread context (8), over-clarify (13), or provide excessive content (8), while QAAgent demands careful follow-up phrasing. 
Thus, LLM agents improve coordination only when interventions match group intent without disrupting conversation.

\subsection{Future Needs}
\label{sec:discussion-future-needs}

\subsubsection{Knowledge Organization}
\label{sec:discussion-knowledge-organization}
Most future requests focus on organizing the shared search process. 
Participants request summaries of conversations or search results (14), outlines or mind maps (13), content marking (7), and duplicate-question handling (5). 
These requests suggest that users need support in maintaining a shared information space, not merely answering queries. 
A useful agent should preserve key findings, connect related evidence, and support revisiting earlier decisions, extending it into a lightweight memory and synthesis layer for collaborative search.

\subsubsection{Search Guidance}
\label{sec:discussion-search-guidance}
Participants also request help deciding what to search next, including refined search suggestions (10), proactively selectable directions (8), user-specified sources (5), and extraction of key content from linked pages (4). 
Some also request task decomposition and work allocation (3), as well as proactive participation in conversations (3).
These requests support guided exploration: the agent should identify gaps, propose alternatives, and distribute work while leaving final decisions to collaborators. 
Moreover, reliable web extraction and synthesis are essential, as poor retrieval can weaken otherwise useful guidance.

\subsubsection{Interaction Control}
\label{sec:discussion-interaction-control}
Future assistance must be easy to invoke, inspect, and control. 
Participants request one-click invocation, copying, and link sharing (9), clearer colors and citations (7), multimodal support for images, video, audio, and files (7), and shared files or click histories (3). 
Interview feedback also warns against slow responses, excessive content, and poorly timed interventions. 
Proactivity should therefore remain controllable: users should be able to trigger, dismiss, revise, or defer summaries and suggestions. 
These requests imply a collaborative workspace with persistent context, transparent evidence, multimodal interaction, and lightweight coordination that preserves user control.

\section{Conclusion}
In this work, we compare LLM-based and conventional support for lightweight collaborative search through a study of 80 participants in 40 pairs. Results indicate that LLM assistance generally reduces perceived task demands and improves collaboration, with effects varying by assistance type. QAAgent supports the most active, responsive conversations and receives the highest collaboration ratings, while CoSearchAgent offers the strongest overall experience and reduces result inspection through source-grounded synthesis. However, users remain concerned about latency, incomplete answers, source credibility, and excessive or poorly timed interventions. Conventional tools produce richer reports than the LLM-based agents, partly because webpages provide more detail than length-limited agent responses. LLM-based reports show less wording overlap across reports. Among scenarios with browsable Web evidence, CoSearchAgent combines the highest report independence with factual richness close to that of the other scenarios. Report quality is also similarly consistent within pairs across scenarios. These findings suggest that collaborative search agents should help teams make sense of shared evidence while leaving users in control of the assistance they receive.










\begin{acks}
This research was sponsored by the Natural
Science Foundation of China (62572475, 61902209, 62377044), Intelligent Social Governance Platform, Major Innovation
Planning Interdisciplinary Platform for the "Double-First Class"
Initiative, Renmin University of China, the Research Funds of Renmin University of China (22XNKJ15),  and Beijing Nova Program.
\end{acks}

\newpage

\bibliographystyle{ACM-Reference-Format}
\balance 
\bibliography{sample-base}

\end{document}